# Quantum improved measurement of time transfer

**ShaoFeng Wang,**[1,2] **Xiao Xiang,**[1,2] **Nicolas Treps,**[3] **Claude Fabre,**[3] **RuiFang Dong,**[1,2,*] **Tao Liu,**[1,2] **ShouGang Zhang**[1,2]

[1]*Key Laboratory of Time and Frequency Primary Standards, National Time Service Center, Chinese Academy of Science, Xi'an, 710600, China*
[2]*School of Astronomy and Space Science，University of Chinese Academy of Sciences, Beijing, 100049, China*
[3]*Laboratoire Kastler Brossel, Universite Pierre et Marie Curie–Paris 6, ENS, CNRS; 4 place Jussieu, 75252 Paris, France*
*\* dongruifang@ntsc.ac.cn*

**Abstract:** Accurate time transfer has become a crucial issue for future space experiments which require increasing resolution over large distances. In 2008, a scheme combining homodyne detection and mode-locked femtosecond lasers was proposed that leads to a potential timing precision reaching the yoctosecond range; with a multimode quantum frequency comb as the input field, the sub-shot noise measurement for the time transfer can further improve the timing precision. Based on this scheme and applying the multimode squeezing frequency comb that was measured to have a phase quadrature quantum noise reduction of 1.5 dB at the analyzing frequency 2MHz, the measurable timing fluctuation was reduced from a shot-noise limited value of 8.9E-23s to 7.5E-23s. To our knowledge, this is the first experimental demonstration of the time transfer measurement that achieves a precision beyond the standard quantum limit (SQL).

## 1. Introduction

Accurate time transfer between remote clocks is fundamental to many areas ranging from navigation, global positioning, electric power generation, to tests of general relativity theory, long baseline interferometry in radio astronomy, gravitational wave observation, and future quantum telecommunications. According to Einstein's theory of time synchronization [1,2], accurate time transfer between two observers A and B is dependent on measuring the arrival times of the incoming light pulses, which consists in repeatedly exchanging light pulses. The conventional method is performed by measuring the arrival time of the maximum of the pulse envelope, which is called incoherent time-of-flight (TOF) measurement [3]. The shot noise limit for the timing accuracy is inversely proportional to the frequency spread $\Delta\omega$, i.e., $(\Delta u)^{tof} \geq (\Delta u)^{tof}_{SQL} = 1/(2\Delta\omega\sqrt{N})$, where $N$ is the total number of photons measured in the experiment during the detection time. With the help of interference between the pulses arriving from A and a local oscillator (LO) derived from the local clock in B, the second method of measuring the arrival time is determined by the phase shift on the signal after traversing a given distance and is referred to as a coherent phase [4,5] (PH) measurement. Such measurement suffers an accuracy which is inversely proportional to the center frequency of the signal $\omega_0$, that is $(\Delta u)^{ph} \geq (\Delta u)^{ph}_{SQL} = 1/(2\omega_0\sqrt{N})$. As $\omega_0 > \Delta\omega$, the phase method has a better ultimate sensitivity than the time-of-flight technique. However, it is limited to relative range changes as the ambiguity range equals half the laser wavelength. Femtosecond optical frequency combs offer an intriguing solution by combining the pulsed nature with the coherence of the carrier. These dual characteristics have found great utility in characterizing the temporal jitter of the sources themselves for applications that include high-fidelity optical time transfer [6]. A new scheme combining femtosecond optical frequency combs and homodyne detection was proposed [7], which shows that a new SQL with a better timing sensitivity of $(\Delta u)_{SQL} = 1/(2\sqrt{N}\sqrt{\omega_0^2 + \Delta\omega^2})$ can be obtained by properly shaping the Local Oscillator (LO) [8-10] in the homodyne detection setup. In addition, according to Cramer–Rao theory, homodyne detection is the optimal strategy for achieving a SQL measurement. Another research [11] shows that one can directly access the desired parameter while being insensitive to fluctuations induced by parameters of the environment such as pressure, temperature, humidity and $CO_2$ content by applying appropriate time shaping on the LO pulse. Recently, an experiment has been accomplished to demonstrate a quantum-limited measurement of the delay between two pulses [12].

Furthermore, this SQL sensitivity can be overcome by using multimode squeezing of frequency combs [13-16]. Benefitting from the large number of photons and from the optimal choice of both the detection strategy and of the quantum resource, the proposed scheme represents a significant potential improvement in space-time positioning [7]. In this letter, we build upon the experimentally generated multimode squeezed light to realize quantum improved measurement of time transfer fluctuation. The applied multimode squeezing frequency comb was generated based on a synchronously pumped optical parametric oscillator (SPOPO), and measured to have a phase quadrature quantum noise reduction of 1.5 dB at the analyzing frequency 2MHz. Through a homodyne detection, the measurable timing fluctuation of the signal field relative to a LO was reduced from a shot-noise limited value of

8.9E-23s to 7.5E-23s. To our knowledge, this is the first experimental demonstration of the time transfer fluctuation that achieves a timing precision beyond the SQL.

## 2. Theoretical model

### 2.1 Quantum optical frequency comb

According to Ref. [13-16] and references therein, SPOPO is an OPO pumped by a train of ultrashort pulses that are synchronized with the pulses making round trips inside the optical cavity. A huge number of pump modes leads to a great complexity of the parametric down-conversion process taking place in the intracavity nonlinear crystal. Through Bloch-Messiah decomposition, a series of eigenmodes of the SPOPO are reduced which are in the shape of the well-known Hermite-Gauss modes in the frequency domain

$$\upsilon_k(u) = \frac{1}{\sqrt{2^n n!}} H_k\left(\frac{u\Delta\omega}{\sqrt{2}}\right) e^{-\frac{(u\Delta\omega)^2}{4}} e^{i\omega_0 u}. \tag{1}$$

These modes are called the supermodes of the SPOPO and form an orthonormal basis of modes. As long as the pump irradiance $P$ is below the SPOPO oscillation threshold $P_{thr}$, these supermodes can be highly squeezed vacuum modes in alternative quadratures. The quadrature component fluctuations of the $k$th supermode injected into a homodyne detection setup can be given as

$$\Delta^2 \hat{P}_k(\Omega) = 1 - \varsigma\eta_{tot} \frac{\gamma_s^2(1+r\Lambda_k/\Lambda_0)^2 - \gamma_s^2(1-r\Lambda_k/\Lambda_0)^2}{\gamma_s^2(1+r\Lambda_k/\Lambda_0)^2 + \Omega^2},$$

$$\Delta^2 \hat{Q}_k(\Omega) = 1 + \varsigma\eta_{tot} \frac{\gamma_s^2(1+r\Lambda_k/\Lambda_0)^2 - \gamma_s^2(1-r\Lambda_k/\Lambda_0)^2}{\gamma_s^2(1-r\Lambda_k/\Lambda_0)^2 + \Omega^2}. \tag{2}$$

Where $\hat{P}_k$ and $\hat{Q}_k$ denote the phase and amplitude quadratures relative to the local oscillator respectively; $\Omega$ is the analysis frequency. $\varsigma$ and $\gamma_s$ are respectively the escape efficiency and decay rate of SPOPO. $\eta_{tot} = \rho\eta\xi^2$ is the total detection efficiency of the homodyne detection setup, where $\rho$ is the quantum efficiency of the balanced photodetectors pair, $\eta$ is the propagation loss of the signal, and $\xi$ is the interference visibility between the signal and LO. $r$ is the normalized amplitude pumping rate and defined as $r = \sqrt{P/P_{thr}}$. $\Lambda_k$ is the eigenvalue of the $k$-th supermode.

### 2.2 Quantum measurements of time transfer variation

Consider the signal is in a fundamental Gaussian form with its mean field described by $\langle E_{(0)}^{(+)}(u) \rangle \propto \sqrt{N}\upsilon_0(u)e^{i\theta_s}$, where $N$ is the mean number of photon and $\theta_s$ a global phase [17]. A small time transfer variation $\Delta u$ of the mean light leads to a modification of the field received in B. As described in Ref. [7], the temporal mode containing the timing signal $\Delta u$ is approximated to $\upsilon_0(u-\Delta u) \approx \upsilon_0(u) + \frac{\Delta u}{u_0} w_1(u)$, where $w_1(u) = \frac{1}{\sqrt{\alpha^2+1}}(i\alpha\upsilon_0(u) + \upsilon_1(u))$ with $\alpha = \frac{\omega_0}{\Delta\omega}$ being called the timing mode, and $u_0 = 1/\sqrt{\omega_0^2 + \Delta\omega^2}$ the normalization factor. Shaping the LO in the same mode of $w_1(u)$ [8-10] and using the balanced homodyne detection scheme, the timing signal $\Delta u$ can be retrieved. The mean signal of the balanced homodyne detection can be given by

$$\langle \hat{D} \rangle \propto 2\sqrt{N \cdot N_{LO}} \left( \frac{\Delta u}{u_0} \cos(\theta_s - \theta_{LO}) + \frac{\alpha}{\sqrt{\alpha^2 + 1}} \sin(\theta_s - \theta_{LO}) \right). \quad (3)$$

Where $N_{LO}$ is the mean number of photon of the local field and $\theta_{LO}$ its phase. When $\theta_s - \theta_{LO} = 0$, the optimal detection is implemented. Under this condition, the variance of the balanced homodyne signal is given by

$$\Delta \hat{D} = \sqrt{\langle \delta^2 \hat{D} \rangle} \propto \sqrt{\frac{N_{LO}}{1+\alpha^2}(\alpha^2 \Delta^2 \hat{P}_0 + \Delta^2 \hat{Q}_1)}. \quad (4)$$

where $\Delta^2 \hat{P}_0$ and $\Delta^2 \hat{Q}_1$ are the phase quadrature fluctuation of the Hermite mode $\upsilon_0(u)$ and amplitude quadrature fluctuation of the Hermite mode $\upsilon_1(u)$, respectively. Assume a signal to noise ratio of 1 $\left( \langle \hat{D} \rangle = \Delta \hat{D} \right)$, the minimum resolvable time variation is thus deduced.

$$(\Delta u)_{min} = \frac{1}{2\sqrt{N}(\omega_0^2 + \Delta \omega^2)} \sqrt{(\omega_0^2 \Delta^2 \hat{P}_0 + \Delta \omega^2 \Delta^2 \hat{Q}_1)}. \quad (5)$$

When $\Delta^2 \hat{P}_0 = \Delta^2 \hat{Q}_1 = 1$, the so-called SQL given by Ref. [8] is then obtained. If the signal field is further squeezed based on the SPOPO, i.e., $\Delta^2 \hat{P}_0 \& \Delta^2 \hat{Q}_1 < 1$, a sensitivity of $(\Delta u)_{min} < (\Delta u)_{SQL}$ can be realized.

### 3. Experimental setup

The experimental setup for the generation and characterization of quantum optical frequency is shown in Fig. 1. We used a commercial Ti: Sapphire laser (Fusion 100-1200, FemtoLasers), which produced nearly Fourier transform-limited pulses with a duration of 130 fs at 815 nm with an average power of 1.4 W and a repetition rate of 75 MHz. A fraction of about 400 mW of the output was directed to a 0.5 mm long BiB3O6 (BIBO) crystal for frequency doubling, and a frequency doubling efficiency of 23% was achieved which was enough for pumping the subsequent SPOPO. About 2.5 mW was used as the local oscillator (LO) for homodyne detection. Besides, and an additional weak seed beam was prepared for parametric phase-sensitive amplification process in the SPOPO. The SPOPO is a 4m-long, multi-folded ring cavity consisting totally 13 mirrors. The cavity was designed singly-resonant for the signal at 815 nm. The reflectivity of the input coupler M1 is $R_{in}$ = 99%. While for the output coupler M10, the reflectivity is $R_{out}$ = 79.5%. All the other mirrors are highly reflective (HR) at 815 nm. The curved mirrors M2 & M13 have a radius of 25 cm, whereas M4 and M11 have a radius of 6 m. A 2 mm long BIBO crystal was put in the characteristic beam waist for SPDC. By comparing the measured cavity finesse, which was about 24, and the theoretical value of 27, the escape efficiency of $\zeta \approx 0.814$ was deduced. Using the Pound-Drever Hall method and feedback on PZT2, the SPOPO cavity length was locked with a locking beam propagating the opposite direction to the signal path. The pump threshold was measured to be $P_{thr} \approx 55\text{mW}$. Below the threshold, phase-sensitive amplification of the seed was observed. The generated multimode squeezed state was then output from the SPOPO and superimposed with the LO at PBS1, the combined optical pulses were split at PBS2 and the resulting two beams were focused onto a pair of PIN photodiodes (Hamamatsu, S3883, quantum efficiency of $\rho = 0.93 \pm 0.02$) for homodyne detection. Using a HWP, the splitting ratio of PBS2 was tuned to balance the homodyne detector. The propagation efficiency of the squeezed state through optical components was carefully evaluated to be $\eta = 0.98 \pm 0.02$. The degree of mode-matching between the LO and the seed pulse out of the SPOPO was optimized to be $\xi=0.89$.

To perform the sub-shot noise measurement of time transfer, a piezoelectric actuator (PZT4) attached to a highly reflective mirror with an angle of incidence (AOI) < 1° was introduced into the signal arm. The modulation of the PZT4 would induce a longitudinal distance change, which is equivalent to a time transfer variation $\Delta u$. Using the method outlined in Ref. [18], the distance change induced by the modulation driven at 2 MHz was calibrated to be 4.96 ± 0.03 pm/V, or 1.65 ±0.01 E-20 s/V in time scale.

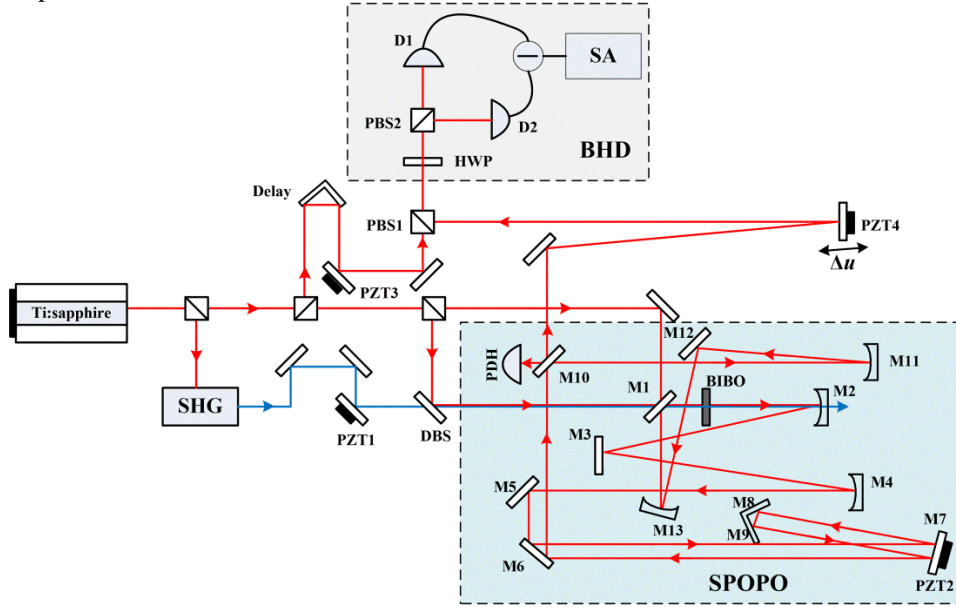

Fig. 1. Set-up for the generation and characterization of quantum optical frequency comb, as well as the measurement of quantum-improved time variation. PBS, polarizing beamsplitter; DBS, dichroic beam splitter; HWP, half-wave plate; PZT, piezoelectric ceramic transducer; SHG, second harmonic generation; M1-M13, cavity mirrors; D1&D2, pin photodetectors; BHD, balanced homodyne detector; SA, spectrum analyzer.

## 4. Experimental results

### 4.1 Generation and measurement of the multimode squeezed state

The first step of the experiment is to characterize the generated squeezing by the SPOPO. By blocking the seed and setting the pump power at 27mW, the squeezed vacuum was measured. The relative phase between the LO and the quantum state was scanned by a sawtooth modulation applied to PZT3 in the LO arm. The measured quadrature variances of the squeezed pulses at 1MHz analyzing frequency were shown in Fig. 2. The black dash line and red solid line are the shot noise and the experimental data, respectively. One can see that, without any loss correction, we got 3 ±0.1 dB of squeezing and 6 ±0.1 dB of anti-squeezing. By using Eq. (2) and experimental parameters, the theoretical calculation was done as well and given by the blue dotted line. The very good agreement between theory and experiments manifest that the cavity round trip loss of SPOPO and the efficiency of detection are the two main restricting factors of improving the generated squeezing.

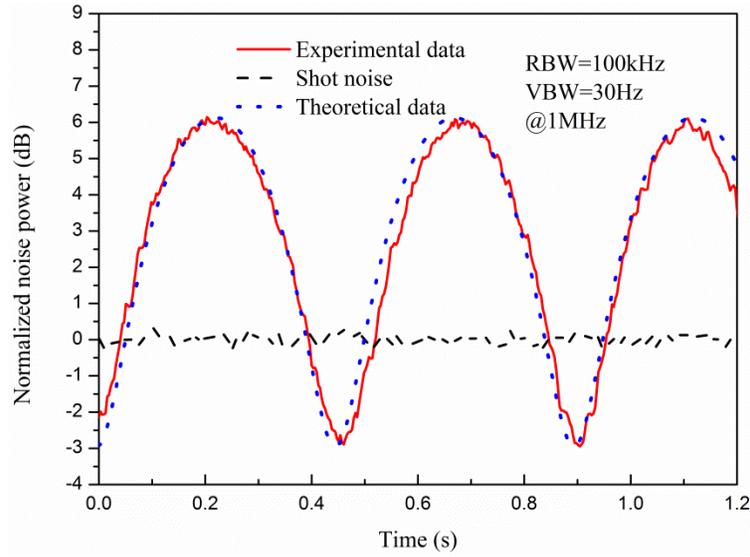

Fig. 2. Measured quadrature variance as a function of the relative phase between the quantum state and the LO. The black dashed line is the shot noise level, the red solid line is the experimental data, and the blue dotted line is the theoretical curve.

According to Eq. (5) [8], the noise fluctuations of the $0^{th}$ and 1st supermodes $\upsilon_0(u)$ and $\upsilon_1(u)$ both contribute to the sensitivity of the achievable $(\Delta u)_{min}$. In our experimental configuration, $\omega_0 \gg \Delta\omega$, $\Delta^2 \hat{Q}_1$ almost has no contribution to $(\Delta u)_{min}$, then the minimum achievable time transfer sensitivity was reduced to $(\Delta u)_{min} = \frac{1}{2\sqrt{N}\omega_0}\sqrt{\Delta^2 \hat{P}_0}$. To demonstrate the sub-shot noise measurement of the time transfer, we need produce the phase-quadrature squeezing in mode $\upsilon_0(u)$ besides maintaining the LO in the mode of $i\upsilon_0(u)$.

By injecting a very weak seed into the SPOPO and locking the relative phase between the seed and the pump to the amplification regime, phase-quadrature squeezing of the output seed in mode $\upsilon_0(u)$ was generated with an optical power of 2 μW. To avoid the presence of excess noise of seed, a larger analyzing frequency of 2 MHz was adopted. Further locking the relative phase between the output seed and LO to π/2, we measured 1.5 ± 0.04 dB of squeezing, which is shown in Fig. 3 by the fine magenta solid line while with the thick black solid line being the shot noise level.

*4.2 Sub-shot noise measurement of the time transfer*

In order to calibrate the time transfer sensitivity, an amplitude modulation was applied on PZT4 for a certain time transfer variation. The minimum detectable time transfer variation occurs for a signal that is equal in magnitude to the background noise, which corresponds to a signal-to-noise ratio of $\Sigma=1$. Due to that the power measured by homodyne is the sum of the time transfer signal power and background noise power, 3 dB improvement of spectrum analyzer when PZT4 is amplitude modulated implied that signal-to-noise ratio of $\Sigma=1$. Firstly 2 μW coherent light produced by blocking the pump was adopted. As seen from Fig. 3, $\Sigma=1$ (3 ±0.05 dB improvement of spectrum analyzer) may be observed from the red dash line when the applied voltage on PZT4 was added to 1.7V, which corresponds to a time transfer variation of $2.8 \pm 0.02\,\text{E}-20\,\text{s}$, thus $(\Delta u)_{min,coh} = 8.9 \pm 0.05\,\text{E}-23\,\text{s}/\sqrt{\text{Hz}}$ was deduced with

RBW of the utilized spectrum analyzer being 100 kHz. Comparing with the SQL calculation based on Eq. 5 ( $(\Delta u)_{SQL}=9.15\text{E}-23\text{s}/\sqrt{\text{Hz}}$ with an optical power of 2 μW and a total detection efficiency of $\eta_{tot}=0.68$ ), a very good agreement was achieved. It verifies that our measurement is shot-noise limited.

Subsequently the similar measurement was accomplished by using the above 1.5dB of phase quadrature squeezed light in mode $\upsilon_0(u)$. From the blue dotted line in Fig. 3, we observed an improvement of the signal-to-noise ratio to $\Sigma_{sqz}=1.183$ (3.8 ± 0.03 dB improvement of spectrum analyzer). Thus the minimum detectable time transfer variation was deduced to be $(\Delta u)_{min,sqz}=7.5\pm0.05\text{E}-23\text{s}/\sqrt{\text{Hz}}$. By comparing the ratio between $(\Delta u)_{min,sqz}$ and $(\Delta u)_{min,coh}$, one can see that such improvement is perfectly equivalent to the squeezing level.

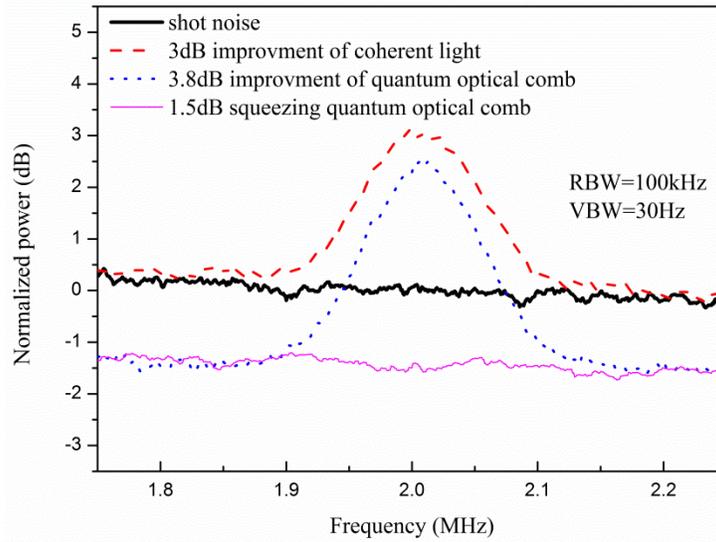

Fig. 3. SNR of quantum optical comb and coherent light with 1.7 V modulation

## 5. Conclusion

Using a 130 fs-scale Ti:sapphire mode-locked laser and SPOPO, we have successfully generated multi-temporal-mode squeezed state. Further applying the generated phase-quadrature squeezing at zero-order supermode, which has a measured squeezing of 1.5 dB at analyzing frequency of 2 MHz, a time variation measurement beyond the standard quantum limit was demonstrated. The result shows that the minimum measurable time variation was reduced from $8.9\pm0.05\text{E}-23\text{s}/\sqrt{\text{Hz}}$ to $7.5\pm0.05\text{E}-23\text{s}/\sqrt{\text{Hz}}$. Such demonstration verifies that quantum optical comb can be used to effectively improve the timing precision beyond the SQL. For an improved squeezing of 10 dB, the precision will reach the ranged of 2.8 $\text{E}-23\text{s}/\sqrt{\text{Hz}}$.

## Funding

National Natural Science Foundation of China (Grant No. 11174282, 91336108, 91636101); Research Equipment Development Project of Chinese Academy of Sciences (Project Name: Quantum Optimization Time Transfer Experiment System Based on Femtosecond Optical Frequency Combs); "Young Top-notch Talents" Program of Organization Department of the

CPC Central Committee (Grant No. [2013]33); Frontier Science Key Research Project of Chinese Academy of Sciences (Grant No. QYZDB-SSWSLH007).